%% file: main.tex
\documentclass[conference]{IEEEtran}
\IEEEoverridecommandlockouts
\usepackage{cite}
\usepackage{amsmath,amssymb,amsfonts}
\usepackage{algorithmic}
\usepackage{graphicx}
\usepackage{textcomp}
\usepackage{xcolor}
\usepackage{array, url}
\usepackage{dblfloatfix}
\usepackage{graphicx, enumitem}
\def\BibTeX{{\rm B\kern-.05em{\sc i\kern-.025em b}\kern-.08em
    T\kern-.1667em\lower.7ex\hbox{E}\kern-.125emX}}
\begin{document}

\title{Towards Trustworthy Edge Intelligence: \\Insights from Voice-Activated Services\\
}

\author{\IEEEauthorblockN{Wiebke Toussaint Hutiri}
\IEEEauthorblockA{\textit{Engineering Systems \& Services} \\
\textit{Delft University of Technology}\\
Delft, The Netherlands \\
0000-0002-9657-9509}
\and
\IEEEauthorblockN{Aaron Yi Ding}
\IEEEauthorblockA{\textit{Engineering Systems \& Services} \\
\textit{Delft University of Technology}\\
Delft, The Netherlands \\
0000-0003-4173-031X}
}

\maketitle

\begin{abstract}
In an age of surveillance capitalism, anchoring the design of emerging smart services in \emph{trustworthiness} is urgent and important. Edge Intelligence, which brings together the fields of AI and Edge computing, is a key enabling technology for smart services. \emph{Trustworthy Edge Intelligence} should thus be a priority research concern. However, determining what makes Edge Intelligence trustworthy is not straight forward. This paper examines requirements for trustworthy Edge Intelligence in a concrete application scenario of voice-activated services. We contribute to deepening the understanding of trustworthiness in the emerging Edge Intelligence domain in three ways: firstly, we propose a unified framing for trustworthy Edge Intelligence that jointly considers trustworthiness attributes of AI and the IoT. Secondly, we present research outputs of a tangible case study in voice-activated services that demonstrates interdependencies between three important trustworthiness attributes: privacy, security and fairness. Thirdly, based on the empirical and analytical findings, we highlight challenges and open questions that present important future research areas for trustworthy Edge Intelligence.
\end{abstract}

\smallskip

\input{sections/1_introduction}

\input{sections/2_background}

\input{sections/3_foundation}
\input{sections/4_voice_services}
\input{sections/5_insights}
\input{sections/6_challenges}

\input{sections/7_conclusion}

\section*{Acknowledgment}
The work was partially supported by the European Union's Horizon 2020 research and innovation programme under grant agreement No. 101021808 and No. 952215.

\bibliographystyle{IEEEtran}
\bibliography{references, manual_references}

\end{document}

%% file: sections/1_introduction.tex
\section{Introduction}

The modern vision of a \emph{smart} world is one in which sensors and devices connected in the Internet of Things (IoT) are augmented with advanced data processing capabilities powered by artificial intelligence (AI). Overlaying this vision with services promises that its power can be harnessed, just as Web services have harnessed the power of the Internet~\cite{Bouguettaya2021internet}. Ultimately, proponents of this vision aspire to create technology that offers fundamental positive change for humanity. However, there is a catch. In a world in which monitoring and monetisation have become the status quo of Web and cloud services, people are increasingly rejecting a future in which their "private human experience [is used] as free raw material for translation into behavioral data”~\cite{zuboff2019age}. Zuboff's exposition of surveillance capitalism, the capture and commodification of personal data for profit-making, is an urgent and compelling wake-up call to reimagine the nature of the emerging \emph{smart} world that we are building as one anchored in \emph{trustworthiness}. 

Edge computing offers a building block for improving the trustworthiness of the computing infrastructure in the IoT-empowered smart world. The Edge enables data processing closer to the source of data collection, which reduces or even eliminates the need to send data to centralised cloud servers \cite{Varghese2021revisiting}. When it comes to user privacy and the protection of personal information, Edge computing can thus fill an important gap. Edge Intelligence broadly encompasses the distribution and execution of AI workloads on and for the Edge~\cite{Deng2020edge}. Edge Intelligence consists of hardware, software, networking and data processing components~\cite{Toussaint2020machine}. Individually these components are already complex technologies. Joined together, interactions between technology layers increase the complexity. Paralleling the complexity of the technology, it is not straight forward to determine what makes Edge Intelligence trustworthy.


This paper scrutinises the requirements for trustworthy Edge Intelligence through the lens of a concrete application scenario of voice-activated services. We contribute to deepening the understanding of trustworthiness in the emerging Edge Intelligence domain in three ways: firstly, we offer a unifying perspective on trustworthy Edge Intelligence that jointly considers trustworthiness attributes of AI and the IoT. Secondly, we present research outputs of a tangible case study that demonstrate interdependencies between three important trustworthiness attributes - privacy, security and fairness - in voice-activated services. Thirdly, based on the findings of our empirical and analytical studies, we highlight future opportunities and challenges for developing trustworthy Edge Intelligence.

We start with a background on trust and trustworthiness in Section~\ref{s:background}. In Section~\ref{s:foundation} we build on the conceptual foundation to align perspectives on trustworthy AI and IoT towards a common vision of trustworthy Edge Intelligence. Section~\ref{s:voice_services} introduces and contextualises voice activation (i.e. technical components that are responsible for enabling and securing access to voice-activated services) within the services ecosystem. We then present insights on trade-offs and interdependencies between privacy, security and fairness attributes in voice-activated services in Section~\ref{s:insights}. In Section~\ref{s:challenges} we take a step back and consider opportunities and challenges in leveraging the insights gained to improve the trustworthiness of voice-activated services in particular, and Edge Intelligence more broadly. Finally, we summarise our work and conclude in Section~\ref{s:conclusion}.

%% file: sections/2_background.tex
\section{Background}
\label{s:background}

Given its constituent technologies, we position that trustworthy Edge Intelligence should at least satisfy the requirements of trustworthy IoT and trustworthy AI. However, trustworthiness concepts in AI and the IoT do not readily align. It is thus not immediately evident what makes Edge Intelligence trustworthy. In this section we present definitions for trust and trustworthiness, and illustrate how trustworthiness is conceptualised in the AI and IoT domains.

\subsection{Trust and Trustworthiness}

Trust and trustworthiness have been studied and formalised in many domains, including AI~\cite{fjeld2020principled}, the Internet of Things~\cite{Yan2014survey}, Cyber Physical Systems~\cite{nist2017framework}, and e-services~\cite{McKnight2002developing}. Drawing on the work of Levi and Stoker~\cite{Levi2000political}, we briefly discuss how we understand trust and trustworthiness in the context of our research. Despite being a contested term, Levi and Stoker position that there is broad consensus across disciplines that trust is relational, seldom unconditional, and a judgement that is expected to inspire a course of action. Trust judgments reflect beliefs about the trustworthiness of the other party. This perspective on trust and trustworthiness is implicitly reflected in services computing, for example conceptualisations of trust in crowd-sourced social IoT, where trust relationships between IoT devices are conditioned on past device performance, which is computed as a reputation score~\cite{Wang2016toward, Nitti2012subjetive}.

Even if trust is not actually required, a trustee (i.e. the party being trusted) can be trustworthy, meaning that they possess the attributes that give a truster (i.e. the party that is trusting) confidence that their trust will not be betrayed. Trustworthiness attributes can be considered along two dimensions: intention and competence. In the eloquent phrasing of Levi and Stoker this means that "the trustworthy will not betray the trust [bestowed upon them] as a consequence of either bad faith or ineptitude." In services computing we assume that services are designed with good intentions and we investigate ill intentions, or bad faith, under the umbrella of security breaches and adversarial attacks (e.g. \cite{Qiu2021adversarial}). The aspects of trustworthiness that relate to intentionality then consider a service's ability to withstand and recover from security breaches and attacks of ill-intentioned actors, rather than the service's own disposition. 

The second dimension of trustworthiness, competence, relates to service attributes that present evidence that the service performs as expected, in alignment with specifications and stakeholder values. In their adaptive trust management framework~\cite{Bahutair2022multiuse}, for example, Bahutair et al. consider two service attributes, security and Internet speed, as trust indicators that are necessary to ensure the free, safe, and secure exchange of IoT services in the absence of a central authority. The trust indicators determine the trustworthiness of the service in the context of its intended usage and in relation to the desired end goal (free, safe and secure exchange of IoT services). 

Having laid a foundation for conceptualising trustworthiness, we now discuss attributes of AI that are deemed necessary to ensure its trustworthiness.

\subsection{Trustworthy AI}

The rapid advancement of AI, accompanied by harmful failures of the technology~\cite{Dobbe2021hard}, has prompted the assembly of trustworthy AI expert groups~\cite{ec2022hlegai}, special interest groups~\cite{ec2022aialliance}, the development of public and private sector AI ethics guidelines~\cite{fjeld2020principled}, and large scale research collaborations to advance the state of trustworthy AI~\cite{tailor2022}. While the understanding of trustworthy AI continues to evolve, key themes are emerging~\cite{fjeld2020principled}. Trustworthy AI attributes that are considered important in the European Union (EU)~\cite{ExpertGroup2019ethics} are summarised in Table~\ref{tab:attributes_tai}. Even though trustworthiness is linked to cultural values and varies across geographic regions, many of the themes in the EU AI Ethics Guidelines are echoed by other guidelines.

\begin{table}[tb]
\begin{centering}
\small
\begin{tabular}{p{0.23\linewidth}p{0.65\linewidth}}  
\textbf{AI attributes} & \textbf{Descriptions} \\ 
\noalign{\smallskip}\hline\noalign{\smallskip}
\textbf{Human agency \& oversight} & Supporting human autonomy and decision making, and promoting a flourishing, democratic and equitable society \\ \noalign{\smallskip}
\textbf{Technical robustness \& safety} & Ensuring physical and mental integrity of humans, and reliable system behaviour that minimises and prevents unintentional, unexpected and unacceptable harm, even under uncertain or adversarial operating conditions \\ \noalign{\smallskip}
\textbf{Privacy \& data governance} &  Protecting the fundamental right to data privacy, including aspects of data quality, integrity, relevance, access and processing\\ \noalign{\smallskip}
\textbf{Transparency} & Communicating system capabilities, purposes and business models openly, making data processing traceable, and decisions explainable so that they can be contested \\ \noalign{\smallskip}
\textbf{Diversity, non-discrimination \& fairness} & Ensuring inclusion and diversity throughout the AI system life cycle, inviting stakeholder participation, and designing for accessibility to ensure equal access and avoid unfair bias \\ \noalign{\smallskip}
\textbf{Societal \& environmental well-being} & Promoting benefit for all human and sentient beings, future generations, society at large, and the environment \\ \noalign{\smallskip}
\textbf{Accountability} & Subjected to scrutiny and redress through auditing and reporting, and consideration of trade-offs posed by trustworthiness concerns\\ \noalign{\smallskip}
\end{tabular}
\end{centering}
\caption{Attributes of Trustworthy AI proposed in the EU AI Ethics Guidelines\cite{ExpertGroup2019ethics}}
\label{tab:attributes_tai}
\vspace{-2.5em}
\end{table}

A central attribute of learning-based AI systems is that their predictive and decision-making capabilities are contingent on data from which the system can learn, and a data-processing pipeline that specifies and performs the learning (typically referred to as model training). This has implications for trustworthy AI. Building on the idea of continuous trust, which states that trust levels can change over time, Toreini et al.~\cite{toreini2020relationship} introduce the notion of a Chain of Trust in machine learning (ML). They argue that the trustworthiness of ML systems should be considered throughout the product lifecycle, and especially at each stage of the ML pipeline. This lifecycle view of trustworthiness is echoed by Suresh and Guttag's~\cite{Suresh2021Framework} framework for identifying sources of harm (broadly referred to as bias) in the ML lifecycle. They illustrate that bias can arise at each stage of the ML lifecycle, and is not only a problem of unrepresentative training data, as is often believed. Bower et al.~\cite{Bower2017Fair} motivate that the fairness attribute of AI trustworthiness should be considered from a pipeline perspective, as compound decisions in ML systems can lead to unfair outcomes, even if individual decisions are fair. Next we discuss how trustworthiness is considered in the IoT.

\subsection{Trustworthy IoT}

Within the Edge Intelligence paradigm, we consider the IoT and Cyber Physical Systems (CPS) from a unified perspective, and jointly refer to them as IoT. This view is motivated by the steady convergence of the two fields, and the benefits of a common perspective which allows us to draw on research progresses in both domains\cite{Greer2019}. Trustworthiness is considered similarly in both fields (see for example the US National Institute of Standards and Technology (NIST) CPS Framework~\cite{nist2017framework} and challenges and opportunities for trustworthy AI published by the Industrial IoT Consortium~\cite{Buchheit2019ai}), and includes attributes (NIST refers to them as \emph{concerns}) of privacy, reliability, resilience, safety and security as described in Table~\ref{tab:attributes_trustworthycps}. These trustworthiness attributes serve to assure that systems behave as expected under various operating conditions. The attributes, while formalised, are viewed as interacting and interdependent, affecting not only each other but also other IoT concerns. 
Interdependencies between attributes raise challenges for trustworthiness, for example the interaction between software and hardware can result in programming bugs that drain the batteries of a critical component, or components developed by different institutions need to be and remain compatible over time\cite{romanovsky2016trustworthy}.

\begin{table}[tb]
\begin{centering}
\small
\begin{tabular}{p{0.23\linewidth}p{0.65\linewidth}}  
\textbf{IoT attributes} & \textbf{Descriptions} \\ 
\noalign{\smallskip}\hline\noalign{\smallskip}
\textbf{Privacy} & Preventing entities from gaining access to data stored in, created by, or transiting the IoT, in order to mitigate risks associated with the processing of personal information\\ \noalign{\smallskip}
\textbf{Reliability} & Delivering stable and predictable performance in expected conditions \\ \noalign{\smallskip}
\textbf{Resilience} & Withstanding instability, unexpected conditions, and gracefully returning to predictable, but possibly degraded, performance\\ \noalign{\smallskip}
\textbf{Safety} & Ensuring the absence of catastrophic consequences on the life, health, property, or data of stakeholders and the physical environment\\ \noalign{\smallskip}
\textbf{Security} & Ensuring that all processes, mechanisms and services are internally or externally protected from unintended and unauthorized access, change, damage, destruction, or use. Considers confidentiality, integrity and availability.\\ \noalign{\smallskip}
\end{tabular}
\end{centering}
\caption{IoT trustworthiness attributes and their definitions from the NIST CPS Framework~\cite{nist2017framework}}
\label{tab:attributes_trustworthycps}
\vspace{-2.5em}
\end{table}

IoT trustworthiness attributes have been studied extensively in Edge Intelligence. For example, on the algorithmic side advances have been made to combine federated learning with local differential privacy to support model training on private, distributed data sources~\cite{truex2020ldpfed}. On the application side, architectures and frameworks that use edge devices for privacy-preserving data stream transformations have been explored for surveillance applications~\cite{Sedlak2022specification}, video analytics~\cite{Lachner2021privacy} and crowd-monitoring~\cite{Stanciu2021privacy}. Hybrid cloud-edge architectures have also been explored for privacy-preserving intelligent personal assistants~\cite{Liang2020paige}. Security attributes have been studied in works like Edgedancer, which presents a platform for portable, provider-independent and secure migration of edge services~\cite{Nieke2021edgedancer}. Having discussed the attributes of trustworthy AI and IoT, we now turn to attributes of trustworthy Edge Intelligence.


%% file: sections/3_foundation.tex
\section{Towards Trustworthy Edge Intelligence}
\label{s:foundation}

In this section we reconcile the AI and IoT perspectives on trustworthiness to gain clarity on attributes that are necessary to ensure trustworthy Edge Intelligence. We first motivate our theoretical foundation for trustworthy Edge Intelligence, and then align trustworthiness attributes between AI and the IoT.

\subsection{Motivation of Theoretical Foundation}
As pointed out by Ding et al. \cite{Ding2021trustworthy}, the truster and trustee in Edge Intelligence can be human, software or cyber-physical objects, like edge hardware and AI models deployed on the edge. In the service computing domain, it is also common that computing tasks are outsourced to different parties, which then become the trustee whose trustworthiness is required. We have pointed out in previous work that neither trustworthy AI attributes, nor trustworthiness concerns in the IoT address the full spectrum of trustworthiness concerns that arise in Edge Intelligence~\cite{Toussaint2020machine}. Using the NIST CPS Framework~\cite{nist2017framework} and the EU AI Ethics Guidelines~\cite{ExpertGroup2019ethics} as a theoretical foundation, we now investigate the alignment between conceptualisations of trustworthy AI and IoT attributes. 

\begin{table*}[hbt]
\begin{centering}
\footnotesize
\begin{tabular}{p{0.14\linewidth}|cccccp{0.41\linewidth}}  
\textbf{Trustworthy} & \multicolumn{5}{c}{\textbf{Trustworthy IoT Attributes}} & \\ 
\textbf{AI Attributes} & \textbf{Privacy}  & \textbf{Reliability}  & \textbf{Resilience}  & \textbf{Safety}  & \textbf{Security} & \textbf{Alignment with Definitions of other IoT concerns}\\ 
\noalign{\smallskip}\hline\noalign{\smallskip}
\textbf{Agency \& Oversight} & & & & & & Manageability, Monitorability, Discoverability, Operability \\ \noalign{\smallskip}
\textbf{Robustness \& Safety} & & X & X & X & X & States, Uncertainty \\ \noalign{\smallskip}
\textbf{Privacy} & X & & & & X & - \\ \noalign{\smallskip}
\textbf{Transparency} & & & & & & Communication, Monitorability, Enterprise, Quality, Utility, Operations on data, Relationship between data, Responsibility, Complexity, Discoverability \\ \noalign{\smallskip}
\textbf{Diversity \& Fairness} & & & & & & Constructivity, Human factors, Usability\\ \noalign{\smallskip}
\textbf{Well-being} & & & & & & Environment\\ \noalign{\smallskip}
\textbf{Accountability} & & & & & & Measurability, Monitorability, Regulatory, Responsibility, Discoverability \\ \noalign{\smallskip}
\end{tabular}
\end{centering}
\caption{Alignment of definitions of Trustworthy AI and IoT attributes (X indicates alignment).}
\label{tab:attributes_edgeai}
\vspace{-2.5em}
\end{table*}

A notable difference between the two frameworks is that the CPS Framework aims to provide a unifying framework that can serve as a reference for the development of CPS tools, standards and documented applications. Concerns (attributes) and descriptions have thus been formulated to support the understanding and development of new and existing CPS, and serve a design purpose within an analytic methodology. It should be noted that trustworthiness is only one of several aspects that is considered in the CPS Framework. 

The EU AI Ethics Guidelines, on the other hand, are driven by ethical and robustness requirements and offer general guidance for building trustworthy AI. While the guidelines aim to provide guidance for operationalising ethical principles for trustworthy AI, they are aspirational in nature, and do not readily convert to concrete design considerations and specifications. Notwithstanding these differences in purpose, we consider the two frameworks a valid starting point for aligning trustworthiness concepts in AI and the IoT.

\subsection{Aligning Trustworthiness Attributes}
The differences between the frameworks are reflected in their descriptions of trustworthiness concepts. In the matrix in Table~\ref{tab:attributes_edgeai} we show which trustworthy AI and IoT attributes align conceptually. \emph{Robustness and safety} in trustworthy AI spans across several trustworthy IoT attributes: the need for reliable system behaviour speaks to \emph{reliability}, performance under uncertain operating conditions relates to \emph{resilience}, minimising and preventing harm translates to \emph{safety} concerns and adversarial operating conditions affect \emph{security}. The \emph{privacy} attribute, on the other hand, is focused on protecting the right to data privacy and the processing of personal information in both domains. In addition, \emph{privacy} in trustworthy AI also includes \emph{data governance}, and aspects of data quality, integrity, relevance and access. \emph{Privacy} in trustworthy AI thus also aligns with the \emph{security} attribute in trustworthy IoT. 

Apart from considering alignment between concepts, it is also worth noting that the same concepts can mean different things in the two domains. For example, the fairness-aware framework for crowdsourcing IoT energy services in~\cite{Lakhdari2021fairness} considers fairness as an optimisation problem, with the goal of maximising the use of energy services across a time period. This perspective diverges from \emph{fairness} in AI, which is concerned with inclusion, diversity, accessibility and bias.  

\subsection{Trustworthiness Interdependencies and Trade-offs}

At first glance, trustworthy AI attributes other than \emph{robustness \& safety} and \emph{privacy} do not overlap with those of trustworthy IoT in Table~\ref{tab:attributes_edgeai}. However, on closer examination the aspirations of trustworthy AI attributes can be mapped to IoT concerns that relate to other (i.e. non-trustworthiness) aspects. To illustrate, the \emph{diversity, non-discrimination \& fairness} attribute of trustworthy AI will influence \emph{human factors} and \emph{usability}, which are part of the \emph{human} aspect in IoT. They also relate to \emph{constructivity}, which is concerned with how the \emph{composition} of modular components satisfies user requirements. Similarly, a lack of \emph{transparency} and \emph{accountability} mechanisms on the side of AI systems will make it difficult for authorised entities to gain and maintain awareness of the state of Edge Intelligence services, thus reducing their \emph{monitorability}.

From Table~\ref{tab:attributes_edgeai} it is clear that to build trustworthy Edge Intelligence, interactions between AI-driven components and traditional IoT components must be considered. Moreover, interdependencies and trade-offs between trustworthiness attributes and other IoT concerns are important, as failures of AI trustworthiness may affect a variety of IoT aspects (e.g. functional, business, composition and human). While this makes intuitive sense, many recent roadmaps and reviews of Edge Intelligence focus only on trustworthy IoT attributes (e.g. \cite{Deng2020edge}, \cite{Merenda2020edge}, \cite{Gill2022ai}). However, this does not mean that the Edge Intelligence community is unaware of the challenges presented by trustworthy AI. Bouguettaya et al. point out that bias and fairness in IoT data analytics are an open problem~\cite{Bouguettaya2021internet} and Ding et al. position the necessity for trustworthy co-design in their roadmap for Edge AI~\cite{Ding2022roadmap}.

In the next sections we present a concrete case study of voice-activated services to illustrate the interdependencies and trade-offs encountered in developing trustworthy Edge Intelligence.

%% file: sections/4_voice_services.tex
\section{Voice Activation in Service Ecosystems}
\label{s:voice_services}

From voice assistants and conversational agents, to social robots and avatars, voice is an important interface for humans to communicate and interact with digital services~\cite{Seaborn2021voice}. Voice assistants such as Apple's Siri, Amazon's Alexa and Microsoft's Cortana have become particularly popular in smart homes, as they enable verbal, hands-free and eye-free interaction with web services (e.g. asking about the weather), personal information (e.g. retrieving calendar information) and smart home devices (e.g. turning on the light). Underlying the seeming simplicity of voice-based interaction lies a complex system of hardware, software, networked communications, machine learning and voice assistant skills. Together with their human and institutional stakeholders, these components constitute the voice-based services ecosystem.

\begin{figure}[hbt]
    \centering
    \includegraphics[width=\linewidth]{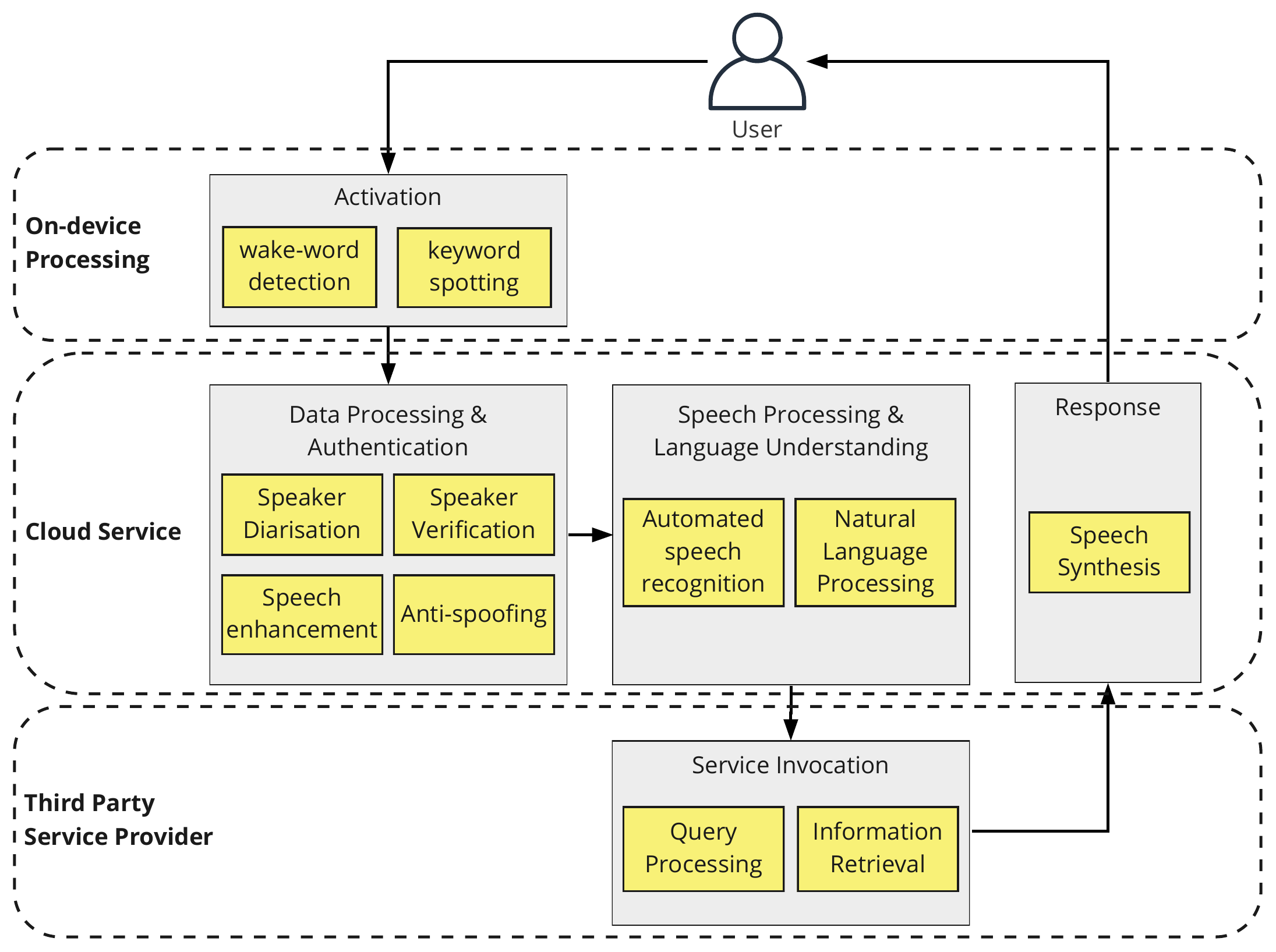}
    \caption{Voice activation and processing in voice assistants}
    \label{fig:voice_processing_steps}
\end{figure}

Figure~\ref{fig:voice_processing_steps} illustrates how technical components are composed for service provision with voice assistants. Data storage and processing tasks are distributed across three layers: at the device level, voice assistants are activated with wake-word detection or keyword spotting on a smart device. Once activated, the device transmits the recorded voice signal to a cloud service provider. Here the voice signal undergoes advanced processing to authenticate and distill the intent of the user. The intent is used to formulate a query, which often invokes a third-party service provider to retrieve the requested information. The query response is sent back to the cloud service provider, which synthesises a spoken response that is transmitted to the device and returned to the user.

We define the voice activation system as the technical components responsible for enabling and securing access to voice-activated services. This includes activation components, namely wake-word detection and keyword spotting, and authentication components, which include speaker diarisation, speech enhancement, speaker verification and anti-spoofing.  Wake-word detection and keyword spotting are examples of on-device machine learning, a type of edge intelligence in which computations are shifted to devices to enhance privacy and reduce latency. Speaker verification is a voice-based biometric that serves an important security function in the system. Anti-spoofing aims to prevent adversarial attacks on speaker verification. Speaker diarisation and speech enhancement are necessary for, but not exclusive to authentication. Together, these components are important as they directly impact whether a user has access to voice-activated services, and if this access is secure and private.

Despite the large-scale adoption of voice-activated services, the current  voice-based ecosystem suffers from weak privacy protection and security vulnerabilities~\cite{Edu2021smart}. We now examine interdependencies between privacy, security and fairness attributes in voice-activated services to illustrate how trade-offs and interactions between them challenge the trustworthy design of Edge Intelligence.




%% file: sections/5_insights.tex
\section{Insights from Voice-Activated Services}
\label{s:insights}

Research into privacy challenges and security vulnerabilities of personal assistant service on smart speakers has revealed several attack surfaces. Edu et al.~\cite{Edu2021smart} categorise security and privacy issues as weak authentication, weak authorisation, profiling, adversarial AI and the complexity of underlying and integrated technologies. While ongoing research efforts have suggested some defenses and mechanisms for addressing the challenges and vulnerabilities, research into this emerging and fast evolving field is still in its early stage. 

\begin{figure}[hbt]
    \centering
    \includegraphics[width=0.6\linewidth]{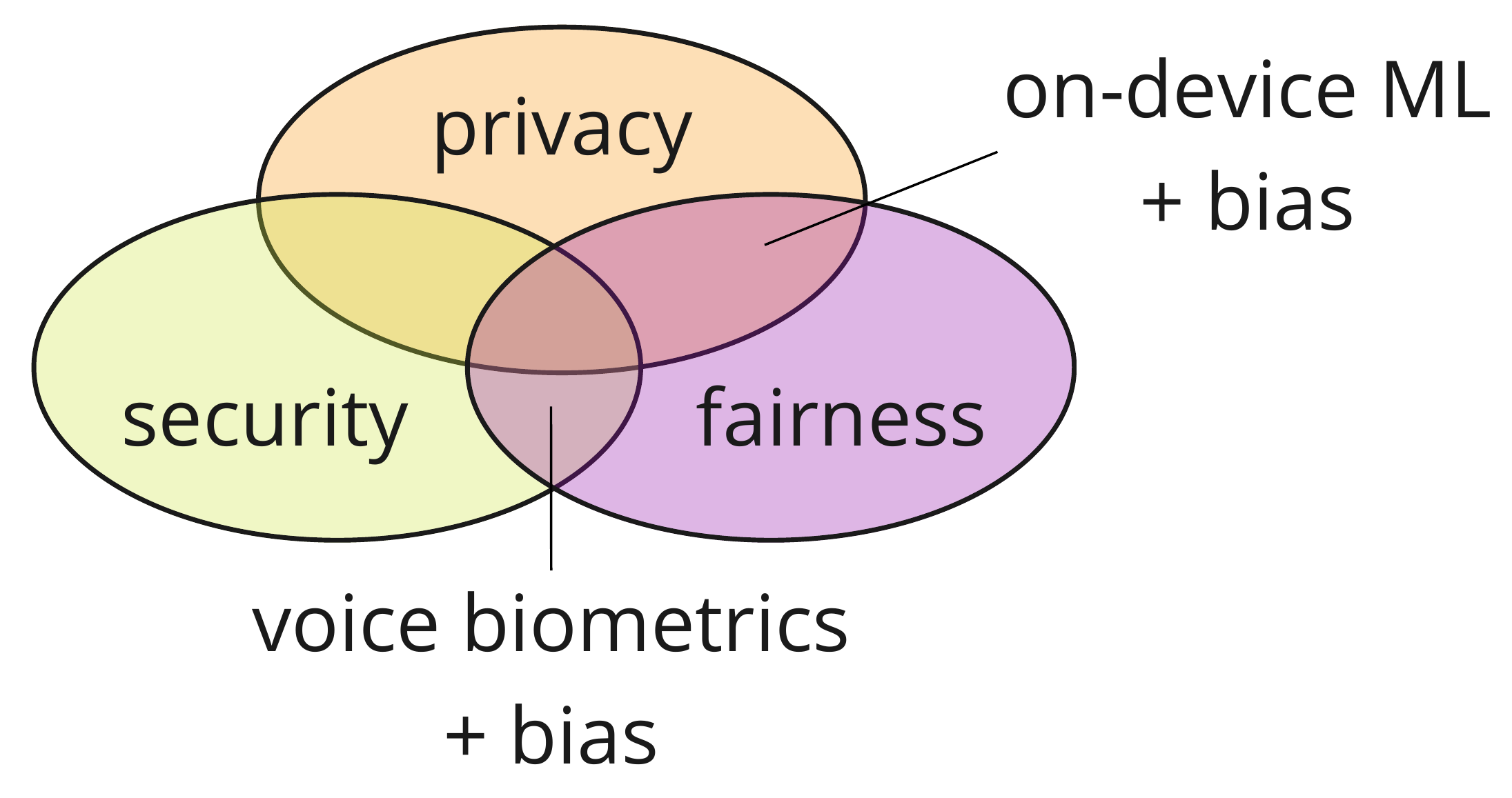}
    \caption{Intersecting trustworthiness concerns in voice-activated services}
    \label{fig:trustworthiness_voice_services}
\end{figure}

In our research we are particularly interested in privacy and security attributes of voice-activated services within the greater context of trustworthy Edge Intelligence. We thus investigated interdependencies and trade-offs between trustworthiness attributes and other system requirements. In this section we highlight results of our recent studies, which provide the first insights into interactions across privacy, security and \emph{fairness}, as illustrated in Figure~\ref{fig:trustworthiness_voice_services}). Specifically, we studied voice biometrics as a defence mechanism for weak authentication, and on-device ML as a solution for enhancing user privacy during inference. We further investigated how bias, a source of unfairness, manifests and propagates through the life cycles of voice service technologies. Our results are clear: \textbf{bias affects the reliability of voice-activated components, and impacts privacy and security attributes}. It should thus be elevated as a first-class trustworthiness consideration alongside security and privacy, to ensure reliable service quality for all users.





\subsection{Impact of Bias on Service Quality and User Experience}

In the AI/ML fairness literature, bias is viewed as a source of unfairness that can result in harm to individual users or even some populations~\cite{Mehrabi2019Survey}. The notion of harm is application dependent and can be considered in different ways~\cite{Barocas2019fairness}. For example, allocative harms are caused when opportunities or resources are withheld from a certain population. Representational harms reinforce stereotypes and subordinate some groups of people along identity lines such as race, class, gender, etc.. When voice activation is biased, this can degrade service quality and user experience for individuals or some groups of people in unpredictable ways. Depending on the third party service that is invoked via voice activation, the consequences may be slight or severe. In addition to degrading service quality and user experience on a service level, bias may also impact user safety. 

One approach to evaluate the service quality of a voice activation system is to consider the system's error rates. While voice activation is a multi-stage process that consists of wake-word detection and speaker verification, the output of the system, from a speaker's point of view, is binary: access is either granted, or denied. If an authorised speaker is denied access, be this because a wake-word is missed or because the speaker's identity could not be validated, this is considered a false negative (FN) error. On the contrary, if an unauthorised speaker is granted access, or if the system is activated erroneously, then this is a false positive (FP) error. FP and FN errors affect different system properties and carry different consequences depending on the third party service that is invoked via voice activation. Table~\ref{tab:harms_consequences} lists system properties, the error type they are affected by, and the consequences of errors. 

\renewcommand{\arraystretch}{1.05}
\begin{table}[bt]
\centering
\footnotesize
\begin{tabular}{p{0.2\linewidth}ccp{0.5\linewidth}}
\textbf{System\newline property} & \multicolumn{2}{c}{\textbf{Affected by}} & \textbf{Consequences of Errors}\\ 
& \textbf{FP} & \textbf{FN} &  \\ 
\hline\noalign{\smallskip}
Usability &  & X & Frustration, feeling ignored, unvalued, excluded \\
Safety &  & X & Injury, disabling injury, loss of life \\
Access &  & X & Denial of access to services \\
Security & X &  & Unauthorised access to personal data and services\\ 
Privacy & X &  & Sensitive information revealed to third parties\\
Compute & X &  & Longer response time, increased power consumption, reduced battery life\\
Data transfer & X &  & Increased financial cost to consumer
\end{tabular} \smallskip
\caption{False positive and false negative errors affect different system properties and carry different consequences for users.}
\label{tab:harms_consequences}
\vspace{-5mm}
\end{table}

In biometric applications intrusion is a particular concern and FP errors pose a security risk, as they grant an unauthorised person access to the system. In device-based applications such as smart speakers and mobile phones, FP errors trigger voice data to be transmitted for downstream processing and thus also affect user privacy, compute and data transfer. Even if no data is transferred, repeated FP errors can increase the compute load on resource constrained devices. Consequences of FP errors can then be that sensitive information is revealed to third parties, that increased compute leads to longer response times and increased power consumption, which again results in reduced battery life, and that users incur increased financial costs due to increased data transfer volumes. 

FN errors, or \emph{misses}, reduce the usability of a device or of downstream voice-activated services. This can leave users feeling frustrated, ignored and excluded. If the downstream services are of critical nature, for example calling medical emergency response, FN errors can also affect user safety and lead to adverse, health-critical consequence. In some applications, like personal identification for banking or social services, FN errors affect access to important services. Being denied access to services can impact users significantly, especially if alternative options to access the services are limited. We now turn to bias in voice biometrics, show how it emerges, and what approaches can be used to mitigate it.


\subsection{Bias in Securing Voice-Authenticated Services}

In Section~\ref{s:voice_services} we illustrated that speaker authentication is necessary for securing voice-activated services from intrusion. Speaker verification systems validate the identity of a person from their voice~\cite{Reynolds2002overview}, which makes them a popular biometric authentication method for securing digital services with voice-based access control. Over the past decade, speaker verification evaluations have shown performance discrepancies between female and male speakers~\cite{Khoury2013speaker}. Historically, these performance differences went uninvestigated, and were attributed to imbalanced training data. While this contributes to bias, imbalanced data offers only a part of the explanation. We conducted a study on bias in automated speaker recognition~\cite{Toussaint2022bias}, where we gathered and analysed empirical and analytical evidence of multiple sources of bias in the well-known VoxCeleb Speaker Recognition Challenge (SRC). Our research shows that historical performance differences between male and female speakers still exist in today's deep neural networks, and that bias is embedded in the development process of speaker verification. 
\smallskip

\subsubsection{Bias in Data Generation}
Even though challenges such as the VoxCeleb SRC serve research purposes and are not necessarily used to evaluate real-life applications, they become benchmarks and shape the research interests and directions of the domain. This makes it a particular concern if they are biased. As expected, we found that bias due to imbalanced representation of speaker groups is one source of bias, with training and evaluation datasets skewed towards males and US nationals. Generated from celebrity speech, the VoxCeleb datasets are also not representative of the broad public. The process of generating the dataset presents additional reasons to raise bias as a concern. Constructed with a fully automated data processing pipeline from open-source audio-visual media, the pipeline directly translated bias that has been exposed in facial recognition verification technology into the speaker verification domain.  
\smallskip

\subsubsection{Bias in Model Building and Implementation}
Beyond bias in the data, we found that modeling choices such as the architecture and feature input can amplify performance disparities. This tends to have a greater negative effect on female speakers and nationalities with fewer speakers. Other sources of bias involve evaluation and engineering practices. Evaluations are based on and optimised for average performance, which hides high error rates for some groups. For example, we found that Indian female speakers have a FP error rate that is 13 times greater than average, indicating that this subgroup is much more exposed to security vulnerabilities than other speakers. Evaluation metrics also introduce bias through normative design decisions such as determining appropriate weights for FP and FN errors. Traditionally, speaker verification has been optimised to reduce security concerns by minimising FP errors. In device-based applications, attributes such as usability, which is influenced by FN errors, are also important. Yet, benchmarks often do not adjust weights to adapt evaluation practices and datasets to these emerging contexts, leading to the oversimplification of common real-life usage scenarios. 
\smallskip

\subsubsection{Mitigating Bias with Inclusive Evaluation Datasets}
High gain approaches to mitigate bias are not limited to algorithmic interventions. We have observed that interdisciplinary approaches to tackle bias with software engineering and design interventions present opportunity for progress in voice-activated services. We have already motivated that evaluation datasets that are representative of usage contexts are particularly important for ensuring unbiased speaker verification performance. To address bias due to unreliable evaluation practices, we thus developed design guidelines for inclusive evaluation datasets that enable robust speaker verification evaluation\cite{Hutiri2022design}. We set up experiments to show that the difficulty grading of data samples in the evaluation set, and the distribution of difficult samples across speakers, have a significant effect on evaluation outcomes. These technical aspects of evaluation datasets were previously not considered in the speaker verification domain. Our experimental results enabled us to make evidence-based suggestions for generating evaluation datasets that are inclusive and also more robust in real-life usage scenarios. 

We now move from security to privacy, discussing how the shift from cloud processing to resource-constrained ML on the Edge affects bias. 


\subsection{Bias in Private Voice Activation}
Edge computing offers opportunities for improved user privacy by processing data locally without transferring it to the cloud. On-device ML inference uses Edge computing to make predictions from sensor data directly on the device that collected it, thus improving the privacy of applications that use personal information. However, the benefit of privacy comes at a cost. Memory, power and storage capacity of devices are constrained, and ML models and computing operations must be adapted to this low resource context. This can affect predictive performance\cite{Dhar2021Ondevice}. Moreover, we found that design choices made to adapt models for on-device inference can also impact bias~\cite{Toussaint2022tiny}. In the following paragraphs we unpack the impact that shifting ML inference tasks from the cloud to devices has on bias in an audio keyword spotting task, and highlight interventions for mitigating bias. 

\smallskip

\subsubsection{Reliability Bias in On-device ML}
Our work is underpinned by the concept of \emph{reliability bias}~\cite{Toussaint2022tiny}. We define reliability bias as disparate on-device ML performance due to demographic attributes of users. In voice-activated services, reliability bias can lead to systematic service failures and consequently disparate service reliability across user groups. Reliability bias can be quantified and evaluated during ML development on an individual or a group level. To illustrate, we consider a ML model as a reliable component for a user group if the group's predictive performance equals the model's overall performance across all groups. If a model performs better or worse than average for a group, we consider it to be biased, showing favour for or prejudice against that group. Both favouritism and prejudice increase reliability bias, though only prejudice reduces the quality of service. It is not possible to favour all groups. If some groups are favoured, there will be other groups that experience prejudice.

\smallskip

\subsubsection{Application Heterogeneity Necessitates Fine-tuning}
Next, we characterised the role of pre-processing parameters in audio-based embedded ML~\cite{Toussaint2021Characterising}. Our studies revealed that decisions pertaining to data input and feature extraction present trade-offs between predictive performance, system efficiency (measured as inference latency) and bias. Moreover, we also found that certain design choices are more robust in uncertain deployment conditions than others. For example, models trained at 16kHz show significant performance degradation when data is sampled at 8kHz after deployment. However, models trained with log Mel spectrogram features are less affected by this change than models trained with MFCC features. These results highlight that tuning pre-processing parameters to meet application requirements, rather than using default parameters for feature extraction, is necessary to ensure that heterogeneous, on-device applications work as intended. 
\smallskip
\subsubsection{Bias due to Design Choices}
We expanded this work to investigate how design choices during ML development impact reliability bias in the on-device setting~\cite{Toussaint2022tiny}. We studied the effects of varying default values of four common design choices: the sensor sampling rate, the model architecture, input features and model pruning, which is used for model compression. We found that models trained at higher sample rates have higher predictive performance and are less biased than those trained at lower sample rates, whereas models trained with smaller architectures tend to be more biased. During post-training optimisation, we found the pruning learning rate to be the hyperparameter with the most significant impact on predictive performance and reliability bias. 

\smallskip

\subsubsection{Mitigating Bias in the On-device ML Workflow}
We can use these insights to make actionable suggestions to help developers navigate the complex on-device ML workflow with fairness in mind: by measuring bias and considering fairness during model selection, parameters can be chosen to train less biased models with only a small cost to predictive performance. Once a set of models has been trained, selecting several models for optimisation, testing a range of optimisation parameters (e.g. pruning hyperparameters) and using a satisficing metric such as reliability bias to consider predictive performance \emph{and} fairness during model selection, help to balance trade-offs between accuracy and bias when applying interventions for model optimisation. Ultimately, careful design holds a lot of opportunities for mitigating bias and deploying fairer models without sacrificing predictive performance of voice-activated services.

%% file: sections/6_challenges.tex
\section{Challenges and Open Questions}
\label{s:challenges}

\begin{figure*}[hbt]
    \centering
    \includegraphics[width=\linewidth]{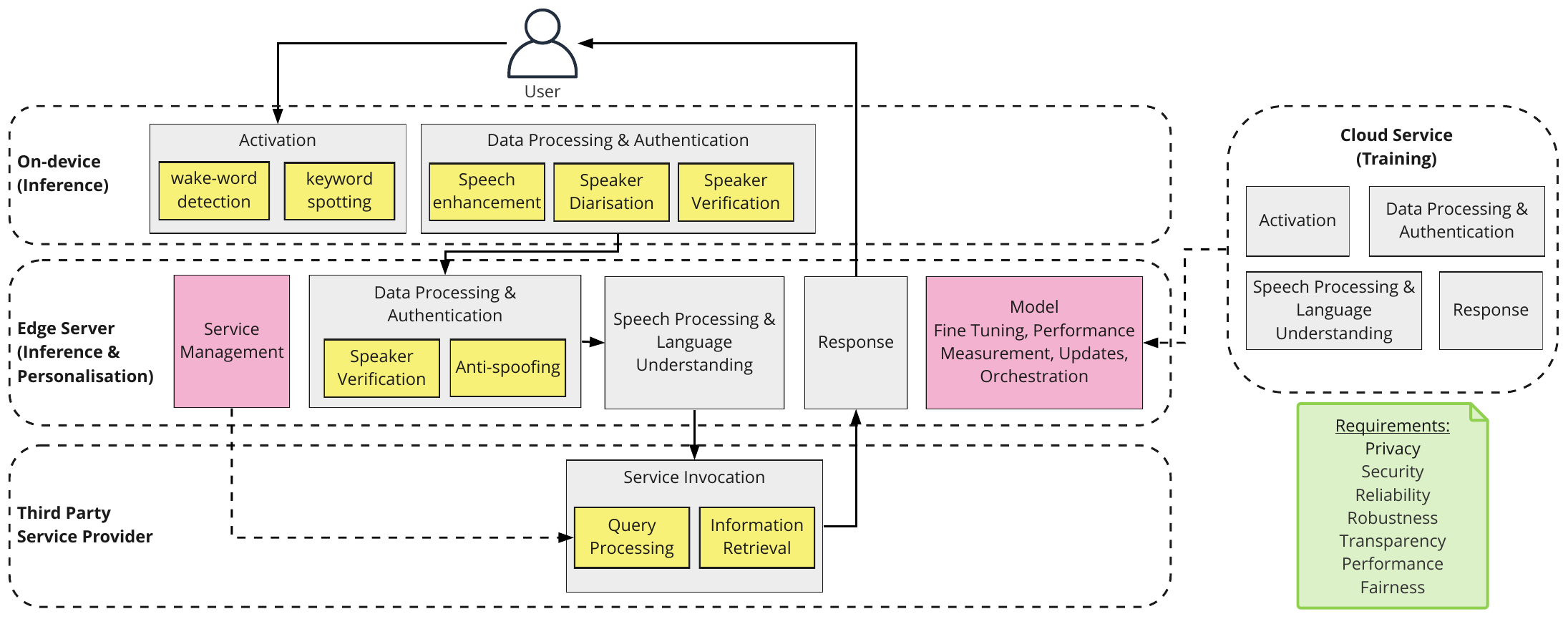}
    \caption{Rethinking technology layers in voice-activated services with trustworthiness in mind}
    \label{fig:edge_supported_voice_processing}
\end{figure*}

In the previous section we looked backwards, highlighting and reflecting on insights that we gained through our research at the intersection of privacy, security and fairness attributes in voice-activated services. In this section we look forward, exploring challenges and open questions that lie ahead on the path towards trustworthy Edge Intelligence. As a source of inspiration we reimagine in Figure~\ref{fig:edge_supported_voice_processing} how technology components and layers could be reassembled in voice-activated services to enforce privacy, improve reliability with personalisation, and encourage the participation of diverse stakeholders.






\subsection{Migrating Inference tasks from Cloud to Edge to Devices}

The current services ecosystem relies heavily on cloud servers for meeting computational demands. While the cloud remains an important computing resource, we need to shift the balance between cloud, Edge and on-device computing to realise our aspirations for trustworthy services. Training large models is unlikely to migrate off the cloud in the short term, but innovations in Edge processing and low resource machine learning make it possible to shift inference, fine-tuning, model updates and management tasks downstream onto Edge servers and devices. An immediate need in voice-activated services is to develop approaches for deploying voice biometrics in on-device low power, low compute settings, in order to secure billions of devices and the services they invoke. Being cognisant of the lessons we learned from on-device keyword spotting, bias should be considered, so that privacy and security do not come at the cost of fairness.

\subsection{Bias Propagation in Voice-Activated Service Composition}

We have discussed bias in two individual components of voice-activated services: keyword spotting and speaker verification. Even though we have investigated interdependencies of trustworthiness attributes, we have not investigated interactions between components. Typically, intelligent systems in smart services are constructed from multiple AI-driven components, as Figure~\ref{fig:edge_supported_voice_processing} shows. Bias does not affect components in isolation, but can propagate through the system, with a high likelihood of touching many components in smart services. For example, two-step cascade architectures are already used for wake-word detection~\cite{gruenstein2017cascade}. The first stage provides an always-on service, and is optimised for extreme energy efficiency and low FN errors. Even though this comes at the cost of an increased FP error rate, the second stage, which runs on a larger processor, can catch the errors downstream. This can reduce performance related bias, but high FP error rates increase the processing load on the second stage, which affects power consumption and battery life. This can lead to different forms of reliability bias pertaining to hardware performance in the second stage of the wake-word spotter. 

Having a more comprehensive understanding of how bias propagates through the system and affects various attributes is thus important for the future. Existing qualitative frameworks can help with this (e.g. the framework proposed by Suresh and Guttag~\cite{Suresh2021Framework}), but new quantitative tools that can be integrated into the development and deployment process are also necessary to facilitate better design.

\subsection{Mitigating Bias with Personalisation and Tolerancing}

Personalisation adapts technology to individual users. This presents a promising avenue for mitigating bias. For example, in speaker verification we found that tuning the classification threshold for groups of same-gender-same-nationality speakers, rather than for all speakers, improves the performance for all groups~\cite{Toussaint2022bias}. A promising direction for future work is to investigate if the same holds true when tuning thresholds for individual users. Further developing algorithmic approaches, like model fine-tuning, for Edge and on-device settings is also promising.

Tolerancing presents an interesting alternative approach for considering ML component performance. While ML is largely concerned with optimising performance, many physical engineering components are designed to a tolerance. Tolerancing implies designing a component to a satisfiable range. Rather than optimising metrics to the highest possible aggregate, ML components that satisfice metrics can aim to meet users' needs and a specified quality of service for all users. The desired outcome are models that perform within an acceptable performance range for all users, rather than particularly well for some, and poorly for others. Tolerancing presents a very different approach to addressing bias, as the end goal is sufficiently good performance for all, rather than equal performance for all. 

Whether personalisation or tolerancing, doing these post-processing operations without compromising user privacy will be important, as parameters such a thresholds contain personal information. Private personalisation may also open new opportunities for human-AI collaboration. An interesting question for future research is whether humans are willing to provide more useful data and feedback to improve system performance if the service is private and they trust it.


\subsection{Trustworthiness Beyond Fairness Beyond Debiasing}

Bias is only one source of unfairness, and fairness only one aspect of trustworthy Edge Intelligence. While developing unbiased Edge Intelligence is a necessary research and design objective, it is also important to study how business models and deployment end goals support diversity and fairness objectives throughout the AI life cycle. For example, if an unbiased model is deployed to monitor and discriminate against a minority group, the outcomes remain unfair~\cite{mozur2019one}. Or if human data labelling~\cite{hao2022how} and content moderation~\cite{perrigo2022inside} practices rely on exploiting workers at best, and violating their human rights at worst, then the models built with these data, even if unbiased, cannot be described as fair. 

Beyond fairness, research questions relating to transparency, accountability and human agency and oversight are largely unexplored in Edge Intelligence. In our pursuit of trustworthy Edge Intelligence, reflecting on these questions can help us gain insights: Can Edge Intelligence be designed to support consumer choice and control? Can systems be designed for flexibility, making AI-driven components interchangeable? What does it mean for the outputs of AI-driven components in Edge Intelligence to be explainable? How is the performance of dynamically evolving Edge Intelligence systems communicated to users, in a way that they can understand and make informed decisions? Who is accountable for the performance of Edge Intelligence; and who is responsible for resolving and repairing issues? As with bias and fairness, attributes like agency and oversight, transparency and accountability interact with each other and with other system components. Many open questions remain, and future research is needed to reveal those interactions, trade-offs and interdependencies of the various trustworthy AI and IoT attributes that enable trustworthy Edge Intelligence.





%% file: sections/7_conclusion.tex
\section{Conclusion}
\label{s:conclusion}

Over the coming years smart services will continue to penetrate our daily lives. As researchers and practitioners, we carry the responsibility of fostering practical processes that create the necessary preconditions to ensure that smart services result in "fundamental positive change for humanity"~\cite{Bouguettaya2021internet}. This paper provides timely insights into the intricacies and opportunities that lie ahead as we move forward on this path towards trustworthy Edge Intelligence. We advocate that trustworthiness is an indispensable requirement when embedding AI on the Edge for advanced IoT services, and that a holistic approach to trustworthiness should inform the growing adoption of Edge Intelligence. By sharing our insights from voice-activated services, we unify trustworthiness perspectives from the AI and IoT domains. Our work highlights that fairness cannot be treated as a retrospective design add-on, but that it should be elevated as a first-class trustworthiness consideration alongside security and privacy. 
